\documentclass[aps,prd,groupedaddress,amssymb,twocolumn,eqsecnum,showpacs,epsfig,nofootinbib]{revtex4}

\usepackage{graphicx}
\usepackage{bm}
\usepackage{dcolumn}
\usepackage{amsmath}

\usepackage{amsmath}
\usepackage{amssymb}

\numberwithin{equation}{section}
\usepackage{epsfig}

\begin{document}
\newcommand{\newc}{\newcommand}

\newc{\be}{\begin{equation}}
\newc{\ee}{\end{equation}}
\newc{\ba}{\begin{eqnarray}}
\newc{\ea}{\end{eqnarray}}
\newc{\bea}{\begin{eqnarray*}}
\newc{\eea}{\end{eqnarray*}}
\newc{\D}{\partial}
\newc{\ie}{{\it i.e.} }
\newc{\eg}{{\it e.g.} }
\newc{\etc}{{\it etc.} }
\newc{\etal}{{\it et al.}}
\newc{\lcdm}{$\Lambda$CDM}
\newcommand{\nn}{\nonumber}
\newc{\ra}{\rightarrow}
\newc{\lra}{\leftrightarrow}
\newc{\lsim}{\buildrel{<}\over{\sim}}
\newc{\gsim}{\buildrel{>}\over{\sim}}
\newcommand{\mincir}{\raise
-3.truept\hbox{\rlap{\hbox{$\sim$}}\raise4.truept\hbox{$<$}\ }}
\newcommand{\magcir}{\raise
-3.truept\hbox{\rlap{\hbox{$\sim$}}\raise4.truept\hbox{$>$}\ }}

\title{Confronting Dark Energy Models using Galaxy Cluster Number Counts}

\author{S. Basilakos}\email{svasil@academyofathens.noa.gr}
\affiliation{Academy of Athens, Research Center for Astronomy and
Applied Mathematics,
 Soranou Efesiou 4, 11527, Athens, Greece}

\author{M. Plionis}\email{mplionis@astro.noa.gr}
\affiliation{Institute of Astronomy \& Astrophysics, Nationals
Observatory of Athens, Thessio 11810, Athens, Greece, and
\\Instituto Nacional de Astrof\'isica, \'Optica y Electr\'onica, 72000 Puebla, Mexico}

\author{J. A. S. Lima}\email{limajas@astro.iag.usp.br}

\affiliation{Departamento de Astronomia (IAGUSP),  Universidade de S\~ao Paulo\\
Rua do Mat\~ao, 1226, 05508-900, S. Paulo, Brazil}

\begin{abstract}
The mass function of cluster-size halos and their redshift distribution
are computed for 12 distinct accelerating cosmological scenarios and
confronted to the predictions of the conventional flat
$\Lambda$CDM model. 
The comparison with $\Lambda$CDM is performed by 
a two-step process. Firstly, we determine the free
parameters of all models through a joint
analysis involving the latest cosmological data, using SNe type Ia,
the CMB shift parameter and BAO. Apart from a brane world inspired
cosmology, it is found  that the derived Hubble relation of the
remaining models reproduce the $\Lambda$CDM results approximately
with the same degree of statistical confidence. Secondly, in order  to
attempt distinguish the different dark energy models from the expectations of $\Lambda$CDM,
we analyze the predicted cluster-size halo
redshift distribution on the basis of two future cluster surveys: (i)
an X-ray survey based on the {\tt eROSITA} satellite, and (ii) a
Sunayev-Zeldovich survey based on the South Pole Telescope. As a
result, we find that the predictions of 8 out of 12 dark energy
models can be clearly distinguished from the $\Lambda$CDM cosmology,
while the predictions of 4 models are
statistically equivalent to those of the $\Lambda$CDM model, as far as the
expected cluster mass function and redshift distribution are
concerned. The present analysis suggest that such a technique appears to
be very competitive to independent tests probing the late time
evolution of the Universe and the associated dark energy effects.

\end{abstract}
\maketitle

\section{Introduction}
Recent studies in observational cosmology lead to the conclusion
that the available high quality cosmological data, from Supernovae
type Ia (SNe Ia), matter power spectrum analysis,  the angular power spectrum
of the cosmic microwave background (CMB), baryon acoustic
oscillations (BAO) and other complementary probes, like the
existence of old galaxies at high redshifts, are well fitted by an
emerging ``standard cosmological model'' (see
\cite{Riess07,Spergel07,essence,Kowal08,komatsu08,Hic09,LJC09,BasPli10}
and references therein). By assuming flatness, as theoretically
required by inflation and the CMB observations, the standard model is
described by the Friedmann equation:

\be H^2(a)=\left(\frac{{\dot
a}}{a}\right)^2=\frac{8 \pi G}{3}\left[\rho_{m}(a)+
\rho_{\rm DE}(a)\right] \;,\label{fe1}
\ee
where $a(t)$ is the scale factor
of the universe, $\rho_{m}(a)$ is the density corresponding to the
sum of baryonic plus cold dark matter (the latter needed to explain
clustering), and an extra component $\rho_{\rm DE}(a)$, with negative
pressure called dark energy (DE hereafter), that accelerates the cosmic
expansion (for reviews see \cite{reviews}).

Nowadays, the physics of dark energy is considered one of the most
fundamental and challenging problems on the interface uniting
Astronomy, Cosmology and Particle Physics. In the last decade there
have been theoretical debates among the cosmologists regarding the
nature of this exotic component. Many candidates have been proposed
in the literature, such as a cosmological constant $\Lambda$
(vacuum), time-varying $\Lambda(t)$ cosmologies, quintessence,
$k-$essence, quartessence, vector fields, phantom, tachyons,
modifications of gravity, Chaplygin gas and the list goes on (see
\cite{Ratra88,Oze87,Weinberg89,Lambdat,Bas09c,Wetterich:1994bg,
Caldwell98,Brax:1999gp,KAM,fein02,Caldwell,Bento03,chime04,Linder2004,
Brookfield:2005td,Grande06,Boehmer:2007qa} and
references therein). Generically, some  high energy field theories
also suggest that the equation of state of such a dark energy is a
function of cosmic time (see, for instance, \cite{Ellis05}).
Naturally, in order to establish  the evolution of its equation of
state (EoS), a realistic form of $H(a)$ is required and must be
constrained through a combination of independent DE probes.

On the other hand, the abundance of collapsed structures as a
function of mass and redshift is a key statistical test for studies
of matter distribution in the universe, and, more importantly, it can
readily be accessed from observations \cite{Evra}. Indeed, the mass
function of galaxy clusters has been measured based on X-ray surveys
\cite{Borg01, Reip02, Vik09}, via weak and strong lensing studies
\cite{Bat98, Dahle06, Corl09}, using optical surveys, like the SDSS
\cite{Bah03, Wen10}, as well as, through Sunayev-Zeldovich (SZ)
effect \cite{Taub05}. In the last decade many authors have been
involved in this kind of studies and have found that the abundance
of the collapsed structures is affected by the presence of a dark
energy component
\cite{Wein03,Liberato,manera,Abramo07,Fran08,Mort09,Pace10,Alam10,Khed10}.

In this work, we discuss how to differentiate a large family of flat
DE cosmologies (12 models) from the conventional
$\Lambda$CDM model. Initially,  a joint statistical analysis,
involving the latest observational data from SNe type Ia, CMB shift
parameter and BAO, is implemented. Since however the resulting 
Hubble functions of these DE models (apart a brane world scenario) are
statistically indistinguishable, we attempt to discriminate the
different DE models by computing the halo mass function
and the corresponding redshift distribution of the cluster-size
halos. Finally, by using future X-ray and SZ surveys
we show that the evolution of the cluster abundances (especially at
large redshifts $z\magcir 0.5$), is a potential discriminator (in the
Popper sense) for a large fraction of the studied DE
models. As an extra bonus, we find that many of the DE models could be
differentiated, from $\Lambda$CDM, even with the present available
observational data from galaxy cluster mass function.

The article is planned as follows. Section 2 is a provision of dark
energy issue and the overall approach adopted in the paper. In
section 3, a joint statistical analysis based on SNe Ia, CMB and BAO
is carried out for the conventional $\Lambda$CDM model. This
statistical approach is used to constrain the DE model parameters
and it is presented in section 4. The linear growth factor of
all models is discussed in section 5, while in section 6, we discuss
and compare the corresponding theoretical predictions regarding the evolution of
the cluster abundances. Finally, the main conclusions are
summarized in section 7. Throughout the paper we adopt
$H_{0}=71$km/sec/Mpc. 

\section{Dark energy Equation of State}

In what follows, it will be assumed that the universe is a
self-gravitating fluid described by general relativity, and endowed
with a spatially flat homogeneous and isotropic geometry.  In
addition, we also consider that it is filled by non-relativistic
matter plus a DE component (or some effective mechanism
that simulates it), and whose equation of state, $p_{\rm
  DE}=w(a)\rho_{\rm DE}$,
is driving the present accelerating stage. Following standard lines,
the Hubble parameter reads:
\begin{equation}
E^{2}(a)=\frac{H^{2}(a)}{H_{0}^{2}}= \Omega_{m}a^{-3}+\Omega_{\rm DE}{\rm
e}^{3\int^{1}_{a} d{\rm lny}[1+w(y)]},  \label{nfe1}
\end{equation}
where $E(a)$ is the normalized Hubble flow, $\Omega_{m}$ is the
dimensionless matter density at the present epoch,
$\Omega_{\rm DE}=1-\Omega_{m}$ denotes the DE density parameter,
and $w(a)$ its EoS parameter. On the other hand, we can express 
the EoS parameter in terms of $E(a)=H(a)/H_{0}$ 
\cite{Saini00} using the Friedmann equations as
\begin{equation}
\label{eos22} w(a)=\frac{-1-\frac{2}{3}a\frac{{d\rm lnE}}{da}}
{1-\Omega_{m}a^{-3}E^{-2}(a)}.
\end{equation}
Since the exact nature of the DE has yet to be found, the
above DE EoS parameter encodes our ignorance regarding the
physical mechanism powering the late time cosmic acceleration.

The methodology described above can also be applied to the framework
of modified gravity (see \cite{Linjen03, Linder2004}). In this case,
instead of using the exact Hubble flow through a modification of the
Friedmann equation one may consider an equivalent Hubble flow
somewhat mimicking  Eq. ($\ref{fe1}$). The key point here is that
the accelerating expansion can be attributed to a kind of
``geometrical'' DE contribution. Now, since the matter
density (baryonic+dark) cannot accelerate the cosmic expansion, we
perform the following parametrization \cite{Linjen03, Linder2004}:
\begin{equation}
E^{2}(a)=\frac{H^{2}(a)}{H_{0}^{2}}= \Omega_{m}a^{-3}+\delta H^{2}.
\label{nfe2}
\end{equation}
Naturally, any modification to the Friedmann equation of general
relativity may be included in the last term of the above expression.
After some algebra one may also derive, using Eqs. (\ref{eos22}) and
(\ref{nfe2}), an effective (``geometrical'') dark energy EoS
parameter, given by:
\begin{equation}
\label{eos222} w(a)=-1-\frac{1}{3}\;\frac{d{\rm ln}\delta
H^{2}}{d{\rm ln}a}.
\end{equation}
The above formulation will be adopted in our statistical analysis
of all DE models discussed in section 4.

\section{Likelihood Analysis: The $\Lambda$CDM case}

In this section we briefly present the basic observational samples
and statistical analysis tools that will be used to constrain the
cosmological parameters of the
DE models. Here we discuss the $\Lambda$CDM model
since it is now widely believed that if a better cosmology is called for,
it will describe a cosmos that looks much like a $\Lambda$CDM model.

(i) Supernovas. In our statistical analysis we consider the
{\em Constitution} set, containing 397 SNe type Ia, as compiled by Hicken et
al. \cite{Hic09}. In order to avoid possible problems related to the
local bulk flow, we use a subset of this sample containing 366 SNe
Ia all with redshifts $z>0.02$. The likelihood estimator is
determined by a $\chi^{2}_{\rm SNIa}$ statistics
\begin{equation}
\label{chi22} \chi^{2}_{\rm SNIa}({\bf p})=\sum_{i=1}^{366} \left[
\frac{ {\cal \mu}^{\rm th} (a_{i},{\bf p})-{\cal \mu}^{\rm
obs}(a_{i}) } {\sigma_{i}} \right]^{2},
\end{equation}
where $a_{i}=(1+z_{i})^{-1}$ is the scale factor of the Universe in
the observed redshift $z_{i}$, ${\cal \mu}$ is the distance modulus
${\cal \mu}=m-M=5{\rm log}d_{L}+25$ and $d_{L}(a,{\bf p})$ is the
luminosity distance, $ d_{L}(a,{\bf p})=c{a}^{-1} \int_{a}^{1}
\frac{{\rm d}y}{y^{2}H(y)} $.

(ii) Baryon Acoustic Oscillations (BAO). In addition to the SNe Ia
data, we also consider the BAO scale produced in the last
scattering surface by the competition between the pressure of the
coupled baryon-photon fluid and gravity. The resulting acoustic
waves leaves (in the course of the evolution) an overdensity
signature at certain length scales of the matter distribution.
Evidence of this excess was recently found in the clustering
properties of the SDSS galaxies \cite{Eis05}, and it provides a
suitable ``standard ruler'' for constraining DE models. In
particular, we consider the following estimator $A({\bf
p})=\frac{\sqrt{\Omega_{m}}}{[z^{2}_{s}E(a_{s})]^{1/3}}
\left[\int_{a_{s}}^{1} \frac{da}{a^{2}E(a)} \right]^{2/3}$, measured
from the SDSS data to be $A=0.469\pm 0.017$, where $z_{s}=0.35$ [or
$a_{s}=(1+z_{s})^{-1}\simeq 0.75$]. Therefore, the corresponding
$\chi^{2}_{\rm BAO}$ function can be written as
\begin{equation}
\chi^{2}_{\rm BAO}({\bf p})=\frac{[A({\bf
p})-0.469]^{2}}{0.017^{2}},
\end{equation}
where ${\bf p}$ is a vector containing the cosmological fitting
parameters.

(iii) CMB Shift Parameter. Another interesting geometrical probe for
dark energy is provided by the angular scale of the sound horizon at
the last scattering surface. It is  encoded in the location
$l_1^{TT}$ of the first peak of the angular (CMB) power spectrum
\cite{Bond:1997wr,Nesseris:2006er}, and may be defined by the
quantity ${\cal R}=\sqrt{\Omega_{m}}\int_{a_{ls}}^1 \frac{da}{a^2 E(a)}$.
The shift parameter measured from the WMAP 7-years data
\cite{komatsu08} is ${\cal R}=1.726\pm 0.019$ at $z_{ls}=1091.36$ [or
$a_{ls}=(1+z_{ls})^{-1}\simeq 9.154\times 10^{-4}$]. In this case,
the $\chi^{2}_{\rm cmb}$ function reads
\begin{equation}
\chi^{2}_{\rm cmb}({\bf p})=\frac{[R({\bf
p})-1.726]^{2}}{0.019^{2}}.
\end{equation}
It should be stressed that for CMB shift parameter, the contribution
of the radiative component, ($\Omega_{R} a^{-4}$, where
$\Omega_{R}\simeq 4.174\times 10^{-5}h^{-2}$) needs also to be
considered \cite{komatsu08}. Note also that the measured CMB shift
parameter is somewhat model dependent but mostly to models which are
not included in our analysis. For example, such is the case when massive
neutrinos are included or when there is a strongly time varying
equation of state parameter. The robustness of the shift parameter
has been tested and discussed in \cite{Elgaroy07}.

In order to put tighter constraints on the corresponding
parameter space of each DE model, the probes described above must be combined
through a joint likelihood analysis\footnote{Likelihoods are
normalized to their maximum values. In the present analysis we
always report $1\sigma$ uncertainties on the fitted parameters. Note
also that the total number of data points used here is
$N_{tot}=368$, while the associated degrees of freedom are: {\em
  dof}$= N_{tot}-n_{\rm fit}$, with $n_{\rm fit}$ the model-dependent number of fitted
parameters.}, given by the product of the individual likelihoods
according to: ${\cal L}_{tot}({\bf p})= {\cal L}_{\rm SNIa}\times
{\cal L}_{\rm BAO} \times {\cal L}_{\rm cmb}$, which translates in the
joint $\chi^2$ function in an addition:
$\chi^{2}_{tot}({\bf p})=\chi^{2}_{\rm SNIa}+\chi^{2}_{\rm
BAO}+\chi^{2}_{\rm cmb}$.
The results of this analysis provide: $\Omega_{m}=0.28\pm 0.01$ with
$\chi_{tot}^{2}(\Omega_{m})/dof\simeq 439.6/367$.

In this concern, it should be remarked that a value of $\chi_{SNIa}^{2}\simeq 439$ is what one gets by using directly the
{\em Constitution} SNIa  set of Hicken et
al. \cite{Hic09} since the number of the
SNIa used (366) dominates the overall $\chi_{tot}^{2}$ 
budget of the joint likelihood. In addition,  the 
corresponding goodness of fit ($\chi_{SNIa}^{2}/dof \simeq 1.2$) is
significantly larger then the one found by Davis et al. \cite{essence}
($\chi_{SNIa}^{2}/dof \simeq 1$) from a previous sample
containing 192 SNIa (see the discussion in \cite{Wei10,Plionis09}).
Such a  discrepancy appears to be the outcome of the different approaches
chosen in order to join the different contributing SNIa
sets. According to Hicken (private communication, 2009) in the case of
the Davis et al. \cite{essence} data the nearby SNIa were imposed to provide
$\chi^{2}/dof \simeq 1$ by hand, while no such fine-tuning was imposed
on the {\em Constitution} set. 
Also the latter set includes distant SNIa which have
typically larger distance modulus uncertainties with respect to those
used in Davis et al. sample \cite{essence}. 
In particular, this means that the higher 
$\chi^{2}_{SNIa}/dof$ value of the {\em Constitution} set should
probably be attributed to
typically lower SNIa distance modulus uncertainties.\footnote{Such a 
possibility has been crudely tested  by Plionis et al. \cite{Plionis09}. A lower value $\chi^{2}_{SNIa}/dof \simeq 1.07$ (similar to that of Davies \cite{essence}) has been obtained by increasing the distance modulus uncertainty of the {\em Constitution} nearby SNIa ($z\le 0.4$) by 20\%.}    

By using the most recent BAO results of Percival et al.
\cite{Perc10}, we have checked and verified that the above
constraints of the flat $\Lambda$CDM model are not appreciably modified. Therefore, the reader should keep
in mind the value $\chi_{tot}^{2}\simeq 439.6$ for any 
further comparison with
results predicted by the alternative DE models.

\section{Constraints on Dark Energy Models}\label{sec.constr}
Let us now present the twelve flat DE models whose free
parameters will be constrained by using the same methodology
and cosmological data as that applied to the $\Lambda$CDM model 
(see previous section).

\subsection{Constant Equation of State (XCDM model)}
In  this kind of cosmology (hereafter XCDM-models) the equation of
state parameter is constant \cite{MT97}. In a point of fact, these
DE models do not have a clear physical motivation. In
particular, for quintessence models driven by a real scalar field, a
constant EoS parameter requires an extreme fine tuning of its
potential and kinetic energy. In spite of that, this subclass of
DE models have been widely used in the literature due to
their simplicity. Notice that DE models with a canonical
kinetic term have $-1\le w<-1/3$. Models with ($w<-1$), sometimes
called phantom dark energy \cite{phantom}, are endowed with a very
exotic nature, like a scalar field with negative kinetic energy.
Now, by using Eq. (\ref{nfe1}) the normalized Hubble parameter
becomes \be E^{2}(a)=\Omega_{m}a^{-3}+(1-\Omega_{m})a^{-3(1+w)}.
\end{equation}

Now, in order to constrain  XCDM-models with the observational data we
sample $\Omega_{m} \in [0.1,1]$ and $w \in [-2,-0.4]$ in steps of
0.01. As a result, we find that the best fit values are
$\Omega_{m}=0.28\pm 0.01$ and $w=-0.99\pm 0.05$ with
$\chi_{tot}^{2}(\Omega_{m},w)/dof \simeq 439.5/366$.

The above results are in excellent agreement with those found by
different authors \cite{essence,Kowal08,{komatsu08},
{Hic09},{Vik09},{BasPli10}}. It is also worth noticing that the
concordance $\Lambda$CDM cosmology is described by a XCDM model with
$w$ strictly equal to -1. The corresponding limits form the basis of
our present comparison and were separately derived (see last
section).

\subsection{Parametric Dark Energy (CPL model)}

This kind of cosmology was first discussed with basis on the
Chevalier-Polarski-Linder \cite{Chevallier:2001qy,Linder:2002et}
parametrization (hereafter CPL). The dark energy EoS parameter is
defined as a first order Taylor expansion around the present epoch:
\begin{equation}
w(a)=w_{0}+w_{1}(1-a),\label{cpldef}
\end{equation}
where $w_{0}$ and $w_{1}$  are constants.  The normalized Hubble
parameter now reads:
\begin{equation}
E^{2}(a)=\Omega_{m}a^{-3}+(1-\Omega_{m})
a^{-3(1+w_{0}+w_{1})}e^{3w_{1}(a-1)}.
\end{equation}
In order to constrain the free parameters,  we sample them as
follows: $w_{0} \in [-2,-0.4]$ and $w_{1} \in [-2.6,2.6]$ in steps
of 0.01. By fixing a prior in the density parameter,
$\Omega_{m}=0.28$, we find that the overall likelihood function
peaks at $w_{0}=-0.96 \pm 0.13$ and $w_{1}=-0.40\pm 0.70$ in very
good agreement with the 7 years WMAP data \cite{komatsu08}. The
corresponding $\chi_{tot}^{2}(w_{0},w_{1})/dof$ is 439.5/366. A
value that should be compared with the $\Lambda$CDM prediction.

\subsection{Braneworld Gravity (BRG model)}
In the context of a braneworld cosmology (hereafter BRG) the
accelerated expansion of the universe can be explained by a
modification of the gravitational interaction in which gravity
itself becomes weak at very large distances (close to the Hubble
scale) due to the fact that our four dimensional brane survives into
an extra dimensional manifold (see \cite{Deff} and references
therein). The interesting aspect of this scenario is that the
corresponding functional form of the normalized Hubble flow as given
by  Eq. (\ref{nfe2}) is affected only by one free parameter,
$\Omega_{m}$. The quantity $\delta H^{2}$ is given by \be \delta
H^{2}=2\Omega_{bw}+2\sqrt{\Omega_{bw}}
\sqrt{\Omega_{m}a^{-3}+\Omega_{bw}} \ee where
$\Omega_{bw}=(1-\Omega_{m})^{2}/4$. From Eq.\ref{eos222}, it is
readily checked that the geometrical DE equation of state
parameter reduces to \be w(a)=-\frac{1}{1+\Omega_{m}(a)} \ee where
$\Omega_{m}(a)\equiv \Omega_{m}a^{-3}/E^{2}(a)$. The joint
likelihood analysis provides a best fit value of $\Omega_{m}=0.23\pm
0.01$, but the fit is much worse,
$\chi_{tot}^{2}(\Omega_{m})/dof\simeq 500/367$, in comparison  with
the one provided by a $\Lambda$CDM cosmology.

\subsection{Low Ricci Dark Energy (LRDE model)}
In this model we use a simple parametrization
for the Ricci scalar which is based on a Taylor expansion around the
present time: ${\cal R}(a)=r_{0}+r_{1}(1-a)$. It is interesting to
mention that at the early epochs the cosmic evolution tends
asymptotically to be matter dominated (for more details see
\cite{Linder2004}). In this framework, we have
\begin{equation}
E^{2}(a)=\left\{ \begin{array}{cc}
        a^{4(r_{0}+r_{1}-1)}{\rm e}^{4r_{1}(1-a)} & \;\;\;\;a\ge a_{t}  \\
       \Omega_{m}a^{-3} & \;\;\;\;a<a_{t}
       \end{array}
        \right.
\label{SS}
\end{equation}
where $a_{t}=1-(1-4r_{0})/4r_{1}$. At present, the matter density
parameter is directly related with the above constants via \be
\Omega_{m}=\left(\frac{ 4r_{0}+4r_{1}-1 } {4r_{1}}
\right)^{4r_{0}+4r_{1}-1} {\rm e}^{1-4r_{0}} . \ee The corresponding
EoS parameter is given by \be w(a)=\frac{1-4{\cal
R}(a)}{3}\left[1-\Omega_{m} {\rm e}^{-\int_{a}^{1}d{\rm lny}(1-4{\cal
R})}\right]^{-1} \;\;. \ee Notice, that we sample the unknown
parameters as follows: $r_{0} \in [0.5,1.5]$ and $r_{1} \in
[-2.4,-0.1]$ and here are the results: $r_{0}=0.80 \pm 0.02$ and
$r_{1}=-0.69 \pm 0.05$ ($\Omega_{m} \simeq 0.27$) with
$\chi_{tot}^{2}(r_{0},r_{1})/dof \simeq 439.9/366$.

\subsection{High Ricci Dark Energy (HRDE model)}
A different Ricci DE model was proposed by Linder
\cite{Linder2007}. In this framework, the Ricci scalar evolves at
high redshifts  obeying the following expression \be {\cal R}\simeq
\frac{1}{4}\left[1-3w_{0} \frac{\delta H^{2}} {H^{2}}\right] \ee
where $\delta H^{2}=E^{2}(a)-\Omega_{m}a^{-3}$. In this 
model the normalized Hubble flow becomes: \be
E^{2}(a)=\Omega_{m}a^{-3} \left(1+\beta a^{-3w_{0}}\right)^ {-{\rm
ln}\Omega_{m}/{\rm ln}(1+\beta)} \ee where
$\beta=\Omega_{m}^{-1}-1$. As in the previous case, the effective
equation of state parameter is again related to $E(a)$ according to
Eq.(\ref{eos22}). Now, by using the same sampling as in the
XCDM-models, we find that the joint likelihood peaks at
$\Omega_{m}=0.28\pm 0.01$ and $w_{0}=-0.99\pm 0.05$ with
$\chi_{tot}^{2}(\Omega_{m},w_{0})/dof \simeq 439.5/366$.

\subsection{Tension of Cosmic Magnetic Fields\\ (TCM model)}
A couple of years ago, Contoupolos and Basilakos  \cite{Cont2007}
proposed a novel idea which is based on the following consideration
(hereafter TCM): if the cosmic magnetic field is generated in
sources (such as galaxy clusters) whose overall dimensions remain
unchanged during the expansion of the Universe, the stretching of
this field by the cosmic expansion generates a tension (negative
pressure) that could possibly explain a small fraction of the DE
 ($\sim 2-5\%$). In this flat cosmological scenario the
normalized Hubble flow becomes: 
\be
E^{2}(a)=\Omega_{m}a^{-3}+
\delta H^{2}, \;\;\;\;\;\delta H^{2}=
\Omega_{\Lambda}+\Omega_{B}a^{-3+n} 
\ee
where $\Omega_{B}$ is the density parameter for the cosmic magnetic
fields and $\Omega_{\Lambda}=1-\Omega_{m}-\Omega_{B}$. From Eq.
(\ref{eos222}) it is readily seen that the effective EoS parameter
which is related to magnetic tension reads \be
w(a)=-\frac{3\Omega_{\Lambda}+n\Omega_{B}a^{-3+n}}
{3(\Omega_{\Lambda}+\Omega_{B}a^{-3+n})}\;\;. \ee  Again, in order
to constrain the parameters,  we sample $\Omega_{B} \in [0.,0.3]$
and $n \in [0,10]$. By considering a prior of $\Omega_{m}=0.28$,
the best fit values are: $\Omega_{B}=0.01\pm 0.005$ and $n=1.80 \pm
0.80$ with $\chi_{tot}^{2}(\Omega_{B},n)/dof\simeq 439.8/366$.

\subsection{Pseudo-Nambu Goldstone Boson\\ (PNGB model)}
In this cosmological model the DE equation of state
parameter is expressed with the aid of the potential $V(\phi)\propto
[1+cos(\phi/F)]$ \cite{Sorbo2007}: \be w(a)=-1+(1+w_{0})a^{F}, \ee
where $w_{0} \in [-2,-0.4]$ and $F \in [0,8]$. We see that for $a\ll
1$, the EoS parameter goes to $w(a)=-1$. Based on this
parametrization the quantity $\delta H^{2}$ (see eq.\ref{eos222}) takes 
the following form 
\be
\delta H^{2}=(1-\Omega_{m})
{\rm exp}\left[\frac{3(1+w_{0})}{F}(1-a^{F}) \right].
\ee 
In this case, our joint statistical analysis yields that the
likelihood function peaks at $w_{0}=-0.97\pm 0.09$ and $F=5.9\pm
3.2$ with $\chi_{tot}^{2}(w_{0},F)/dof\simeq 439.7/366$.

\subsection{Early Dark Energy (EDE model)}
Another cosmological scenario that we include in our paper is the
early dark energy model (hereafter EDE). The basic assumption here
is that at early epochs the amount of DE is not negligible
\cite{Doran2006}. In this framework, the overall DE
component is given by 
\be
\Omega_{\rm DE}(a)=\frac{1-\Omega_{m}-\Omega_{e} (1-a^{-3w_{0}})}
{1-\Omega_{m}+\Omega_{m}a^{3w_{0}}}+\Omega_{e}(1-a^{-3w_{0}}) 
\ee
where $\Omega_{e}$ is the early DE density and $w_{0}$ is
the equation of state parameter at the present epoch. We observe
that the EDE model was designed to satisfy the two following
 conditions: (i) mimic the effects of a late dark
energy, and (ii) provide a decelerating expansion of the Universe at
high redshifts. The normalized Hubble parameter is now written as
\be E^{2}(a)=\frac{\Omega_{m}a^{-3}}{1-\Omega_{\rm DE}(a)}, \ee  and from
Eq. (\ref{eos22}), one may obtain the EoS parameter as a function of
the scale factor.

Now, from the joint likelihood analysis we find that
$\Omega_{e}=0.02\pm 0.02$ and $w_{0}=-1.04^{+0.07}_{-0.11}$ (for the
prior of $\Omega_{m}=0.28$) with
$\chi_{tot}^{2}(\Omega_{e},w_{0})/dof\simeq 439.3/366$. Note, that a
different class of EDE models was recently studied in \cite{Alam10}.

\subsection{Variable Chaplygin Gas (VCG model)}
Accelerating cosmologies can also be driven by the so called
variable Chaplygin gas (hereafter VCG) which corresponds to a
Born-Infeld tachyon action \cite{Bento03,Guo05}. Recently, an
interesting family of Chaplygin gas models was found to be
consistent with the current observational data \cite{Vcgdata}. In
the framework of a spatially flat geometry, it can be shown that the
normalized Hubble function takes the following formula:
\begin{equation}
E^{2}(a)=\Omega_{b}a^{-3}+(1-\Omega_{b})
\sqrt{B_{s}a^{-6}+(1-B_{s})a^{-n}}
\end{equation}
where $\Omega_{b}\simeq 0.0226h^{-2}$ is the density parameter for
the baryonic matter \cite{komatsu08} and $B_{s} \in [0.01,0.51]$ and
$n\in [-4,4]$. The effective matter density parameter is:
$\Omega_{m}=\Omega_{b}+(1-\Omega_{b})\sqrt{B_{s}}$. We find that the
best fit parameters are $B_{s}=0.05\pm 0.01$ and
$n=1.65^{+0.19}_{-0.25}$ ($\Omega_{m}\simeq 0.26$) with
$\chi_{tot}^{2}(B_{s},n)/dof \simeq 441.3/366$.

\subsection{Time Varying Vacuum ($\Lambda_{RG}$
and $\Lambda_{PS}$ models)} Let us now consider the possibility of a
decaying $\Lambda$-cosmology, that is, $\Lambda=\Lambda(a)$. In this
kind of  model, the value of $\Lambda$ is small because the Universe
is too old, and, therefore, it alleviates the cosmological constant
problem.  As it appears,  the decaying vacuum equation of state
does not depend on whether $\Lambda$ is strictly constant or
variable. Therefore, the EoS takes the usual form,
$P_{\Lambda}(t)=-\rho_{\Lambda}(t)=-\Lambda(t)/8\pi G$. The global dynamics
of such models have been investigated extensively in the literature,
in fact, much before the discovery of the present accelerating stage
\cite{Lambdat} (for a recent and extensive discussion see Basilakos
et al. \cite {Bas09c}). By introducing in the Friedman equations the
idea of a time-dependent vacuum one obtains
\begin{equation}
 \label{frie34}
\dot{H}+\frac{3}{2} H^{2}=\frac{\Lambda}{2}\,.
\end{equation}
The traditional $\Lambda=const$ cosmology can be described directly
by the integration of Eq. (\ref{frie34}), but the same equation is
also valid for $\Lambda=\Lambda(a)$. This means that a supplementary
equation for the time evolution of $\Lambda$ is needed in order to
unveil the dynamics of this pattern. In this work we consider two
different versions of the $\Lambda(a)$ models, namely
renormalization group of the quantum field vacuum
\cite{Shap00,Grande06} (hereafter $\Lambda_{RG}$) and a power series
vacuum \cite{Bas09b}  (hereafter $\Lambda_{PS}$)
\begin{equation}
\Lambda(H)=\left\{ \begin{array}{cc} \Lambda_0+
3\gamma\,(H^{2}-H_0^2)&
       \mbox{for $\Lambda_{RG}$}\\
H_{0}(\gamma-3\Omega_{m})H+(3-\gamma)H^{2} & \mbox{for
$\Lambda_{PS}$}
       \end{array}
        \right.
\end{equation}
Naturally, the vacuum energy density is normalized to its present
value: $\Lambda_0\equiv\Lambda(H_0)=3\Omega_{\Lambda}H^{2}_{0}$
($\Omega_{\Lambda}=1-\Omega_{m}$). Inserting the above expressions
into Eq. (\ref{frie34}) we finally obtain the normalized Hubble flow
\begin{equation}
E(a)=\left\{ \begin{array}{cc}
\left[\frac{1-\Omega_{m}-\gamma}{1-\gamma}+\frac{\Omega_{m}}
{1-\gamma}a^{-3\,(1-\gamma)}\right]^{1/2} &
       \mbox{for $\Lambda_{RG}$}\\
1-\frac{3\Omega_{m}}{\gamma}+
\frac{3\Omega_{m}}{\gamma}a^{-\gamma/2} & \mbox{for $\Lambda_{PS}$}
       \end{array}
        \right.
\end{equation}
It has been shown \cite{Shap00},\cite{RGTypeIa} that for
$\Lambda(t)$ there is a coupling between the time-dependent vacuum
and matter component. Indeed, by combining the conservation of the
total energy with the variation of the vacuum energy, one can shown
that the $\Lambda(t)$ models provide either a particle production
process or that the mass of the dark matter particles increases
\cite{AL05}.

Now, by applying our statistical procedure for both cosmologies,
the best fit parameters are:\\

$\Lambda_{RG}$  model: $\Omega_{m}=0.28\pm 0.01$ and $\gamma=0.002
\pm 0.001$ with $\chi_{tot}^{2}(\Omega_{m},\gamma)/dof \simeq
439.5/366$.\\

$\Lambda_{PS}$ model: $\Omega_{m}=0.32\pm 0.02$ and
$\gamma=3.45^{+0.02}_{-0.03}$ with
$\chi_{tot}^{2}(\Omega_{m},\gamma)/dof \simeq 440.7/366$.

\subsection{Creation Cold Dark Matter (CCDM Model)}
In the context of the standard general relativity theory, is also
possible to reduce the dark sector by considering the presence of
the gravitationally induced particle creation mechanism. Like
$\Lambda$(t)-models this kind of scenario has also been discussed
long ago \cite{CCDM1}. More recently, the cosmological equations for
the mixture of radiation, baryons and cold dark matter (with
creation of dark matter particles), and the energy conservation laws
for each component have been investigated thoroughly by
\cite{LSS08,SSL09,LJO09}. In this framework, the basic equation
governing the global dynamics of a Universe endowed with  a  flat
geometry is \cite{LSS08,SSL09,LJO09,BasLim} \be \label{CCDM}
\dot{H}+\frac{3}{2}H^{2}=\frac{4\pi
G\rho_{dm}}{3}\;\frac{\Gamma}{H}, \ee where $\rho_{dm}$ is the dark
matter energy densities and $\Gamma$ is the so called creation rate
of the cold dark matter and it has units of $(time)^{-1}$. The
creation pressure is negative and defined in terms of the creation
rate and other cosmological quantities by the expression
\cite{LSS08,SSL09,LJO09,BasLim}
\begin{equation}
\label{CP}
    p_{c} = -\frac{\rho_{dm} \Gamma}{3H}.
\end{equation}
In  CCDM models, the functional form of $\Gamma$ is
phenomenologically parametrized \cite{LJO09} by the following
relation
\begin{equation}
\label{PS2}
\Gamma=3\tilde{\Omega}_{\Lambda}\left(\frac{\rho_{co}}{\rho_{dm}}\right)H,
\end{equation}
where $\tilde{\Omega}_{\Lambda}$ (called $\alpha$ in the
\cite{LJO09}) is a constant, $\rho_{co}=3H^{2}_{0}/8\pi G$ is the
present day value of the critical density. Performing now the
integration of Eq. (\ref{CCDM}) we obtain
\begin{eqnarray}
\label{hub1} E^{2}(a)=\tilde{\Omega}_{m}a^{-3} +
\tilde{\Omega}_{\Lambda}
\end{eqnarray}
where $\tilde{\Omega}_{\Lambda}=1-\tilde{\Omega}_{m}$.
Interestingly, the CCDM model resembles the global dynamics of the
concordance $\Lambda$CDM cosmology by including only one free
parameter. In particular, this means that
$\tilde{\Omega}_{m}=0.28\pm 0.01$ with
$\chi_{tot}^{2}(\tilde{\Omega}_{m})/dof\simeq 439.6/367$.

\section{Evolution of matter perturbations}
In this section we generalize
the basic equations which govern the behavior of the matter
perturbations within the framework of the previously described DE 
models. We focus our attention to the most generic models, ie., the
$\Lambda(t)$ and CCDM models. In these cases, a neo-Newtonian
description  can be introduced by considering an extended continuity
equation together with the Euler and Poisson equations
\cite{Arc94,LZB97}. In virtue of the particle creation model, the evolution
equation of the growth factor becomes \cite{Bas09c,BasLim}:

\begin{equation}
\label{eq:11} \ddot{D}+(2H+Q)\dot{D}-\left(4\pi {\tilde G} \rho_{m}
-2HQ-\dot{Q} \right)D=0,
\end{equation}
where $\rho_{m}$ is the matter density, ${\tilde G}=G$ and
\begin{equation}
Q(t)=\left\{ \begin{array}{cc} -\dot{\Lambda}/\rho_{m}
\;\;\;\;\;\;\;\;\;\;\;\;\;\;\;\;\;
       &\mbox{for $\Lambda_{RG}$ or $\Lambda_{PS}$}\\
3\tilde{\Omega}_{\Lambda}\rho_{co}H/\rho_{m} & \mbox{for CCDM}
       \end{array}
        \right.
\end{equation}

In the case of non interacting DE models, [$Q(t)=0$, ${\tilde G}=G$], the
above equation (\ref{eq:11}) reduces to the usual time evolution
equation for the mass density contrast
\cite{Peeb93,Linjen03,Stab06}, while in the case of the geometrical DE 
models [$Q(t)=0$, ${\tilde G}=G_{\rm eff}$] (eg.,
\cite{Gann09}).

Useful expressions of the growth
factor have been given by Peebles \cite{Peeb93} for the
$\Lambda$CDM cosmology. Several works have also derived the growth factor for
$w=$const DE models 
\cite{Silv94,Wang98,Bas03,Nes08}, and for the braneworld cosmology \cite{Lue04}.
Linder \cite{Linder2004} and Linder and Cahn \cite{Linca08} derived
similar expressions for dark
energy models in which the Ricci scalar varies with time (and which we
use in the current study),
as well as for models with a time varying equation of state, while for the
scalar tensor models the growth factor was obtained by Gannouji \& Polarski
\cite{Gann08}. For the interacting DE models we use
the growth factors which were recently derived in 
\cite{Abramo07,Bas09c,BasLim}.\footnote{For many DE models, it 
is convenient to study the
growth evolution in terms of the expansion scale $a$ or
characteristic scale ${\rm ln} a$, rather than $t$. 
As an example, in the case of the braneworld
  cosmology, Linder \& Cahn \cite{Linder2007} used 
$g=d{\rm ln} (D/a)/d{\rm ln}a$ (see their eqs. 8, 20
  and 27), while for the LRDE and HRDE models, 
Linder \cite{Linca08} utilized $g=D/a$ (see eq. 31 in his paper). 
Since the pure matter universe (Einstein de-Sitter) has the solution of
$D_{\rm EdS}=a$, we normalize our DE models such as to get $D\simeq a$ at
large redshifts due to the dominance of 
the non-relativistic matter component.
(e.g.,
$z=30$, where the DE density is almost negligible and 
the radiation density is less than $\sim 1\%$ of the matter density).}

\begin{table*}[h]
\tabcolsep 5pt 
\vspace {0.2cm}
\begin{tabular}{lccc|ccc|ccc|} \hline \hline
Model     & symbol & $\sigma_{\rm 8, DE}$ & $\delta_c$ & \multicolumn{3}{|c}{($\delta{\cal N}/{\cal
N}_{\Lambda})_{\rm eROSITA}$} &
\multicolumn{3}{|c|}{$(\delta {\cal N}/{\cal N}_{\Lambda})_{\rm SPT}$} \\
          &  &  &   &  $z<0.5$ & $0.5\le z <1$ & $1\le z <1.5$
                    &  $z<0.5$ & $0.5\le z <1$ & $1\le z <1.5$ \\ \hline
XCDM      & black thin line   & 0.802 & 1.675 &-0.01 & -0.01 & -0.01$\pm 0.07$ &-0.01 & -0.01  & -0.01 \\
{\bf BRG} & blue short dashed & 0.523 & 1.667 &-0.95 & -1.00 & -1.00 & -0.95   &-0.99 & -1.00 \\
{\bf CPL} & green dashed line & 0.821 & 1.663 & 0.12 &  0.22 &  0.48$\pm 0.09$ & 0.12 & 0.20& 0.29 \\ 
{\bf LRDE}& red x             & 0.784 & 1.674 &-0.14 & -0.22 & -0.36$\pm0.05$  &-0.14 & -0.19& -0.25 \\
HRDE      & cyan stars        & 0.798 & 1.674 &-0.01 & -0.01 &  0.00$\pm 0.07$ &-0.01 & -0.01 & 0.00 \\
TCM       & magenta dashed    & 0.799 & 1.674 &-0.02 & -0.04 & -0.07$\pm 0.06$&-0.02 & -0.03 & -0.04 \\
PNGB      & red small dots    & 0.802 & 1.674 &0.00  &  0.00 & 0.02$\pm$0.07 & 0.00 & 0.01& 0.01\\
{\bf EDE} & blue long-dashed line & 0.755 & 1.672 &-0.23 & -0.45 & -0.75$\pm 0.03$ & -0.23   & -0.39 & -0.57 \\
{\bf VCG} & black squares     & 0.764 & 1.642 &-0.23 & -0.35 & -0.56$\pm$0.04&-0.22 & -0.30& -0.44\\
${\bf \Lambda_{PS}}$ &red circles  & 1.017 & 1.675 & 2.18  & 22.8 & 3661$\pm 174$  & 2.12 & 11.8 & 119\\
${\bf \Lambda_{RG}}$ &black thick line & 0.795 & 1.675 &-0.05 & -0.11& -0.26$\pm$0.05  & -0.05 & -0.10& -0.16\\
{\bf CCDM}& magenta stars     & 0.564 & 1.675 &-0.60& -0.73& -0.97$\pm$0.1 & -0.58& -0.670& -0.91\\ \hline
\end{tabular}
\caption[]{Numerical results. The $1^{st}$ column indicates the DE
 model. $2^{nd}$ column are symbols or
line types of the models appearing in Figs. 2, 3 and 4. $3^{rd}$ and $4^{th}$
columns show the $\sigma_{\rm 8, DE}$ and $\delta_c$ values.
The remaining columns present the fractional relative difference between the
models and  the $\Lambda$CDM cosmology for two future
cluster surveys discussed in the text. Bold letters denote those
models that can be clearly distinguished from flat $\Lambda$CDM model
at some redshift bin (uncertainties appear only when their value is $\ge 10^{-2}$).}
\end{table*}

\section{Halo abundances and their Evolution in Dark Energy Models}

In this section we derive the cluster-size halo number counts and their
corresponding redshift distribution within the framework of the DE
models analyzed in this work. We will then compare our predictions
with those of the
conventional $\Lambda$CDM cosmology. In principle, this can help us
to understand better the theoretical expectations of the current
DE models, as well as to identify the realistic variants of
the concordance $\Lambda$CDM cosmology.

An important theoretical question in cosmology is how to determine the fraction of
matter in the universe that has formed bounded structures, and what is its distribution
in mass at any given redshift after recombination. The simplest successful
answer to this question was given in 1974  by Press and Schecther \cite{press} (hereafter PSc).
In their approach, the primordial density fluctuations for a given
mass M on the dark matter fluid is described by a random Gaussian
field.  The function, $F(M,z)$,  represents the fraction of the
universe that has collapsed by the redshift $z$ in halos above some
mass $M$. With this  function  one may estimate the number density of halos, $n(M,z)$,
with masses within the range $(M, M+\delta M)$: \be n(M,z) dM=
\frac{\partial {\cal F}(M,z)}{\partial M} \frac{{\bar \rho}}{M}
dM. \ee 
Performing the differentiation and after some algebra one obtains
the following:
\begin{eqnarray}
n(M,z) dM &=& -\frac{\bar{\rho}}{M} \left(\frac{1}{\sigma} \frac{d
\sigma}{d M}\right)
f_{\rm PSc}(\sigma) dM \nonumber\\
&=& \frac{\bar{\rho}}{M} \frac{d{\rm \ln}\sigma^{-1}}{dM} f_{\rm
PSc}(\sigma) dM,
\end{eqnarray}
where $f_{\rm PSc}(\sigma)=\sqrt{2/\pi} (\delta_c/\sigma)
\exp(-\delta_c^2/2\sigma^2)$, $\delta_{c}$ is the 
linearly extrapolated density
threshold above which structures collapse \cite{eke}, while
$\sigma^2(M,z)$ is the mass variance of
the smoothed linear density field, extrapolated to redshift $z$ at which
the halos are identified. It is given in Fourier space by:
\be
\label{sig88}
\sigma^2(M,z)=\frac{D^2(z)}{2\pi^2} \int_0^\infty k^2 P(k) W^2(kR) dk \;,
\ee
where $D(z)$ is the growth factor of perturbations, $P(k)$ is the
power-spectrum of the linear density field, $W(kR)=3({\rm
  sin}kR-kR{\rm cos}kR)/(kR)^{3}$ is the top-hat smoothing function
which contains on average a mass $M$ within a radius 
$R=(3M/ 4\pi \bar{\rho})^{1/3}$ and 
$\bar{\rho}=2.78 \times 10^{11}\Omega_{m}h^{2}M_{\odot}$Mpc$^{-3}$. 
We use the CDM power spectrum, $P(k)=P_{0} k^{n} T^{2}(\Omega_{m},k)$, 
with
$T(\Omega_{m},k)$ the CDM transfer function (according to
\cite{Bard86}), $n\simeq 0.96$, following the 7-year WMAP results 
\cite{komatsu08}, and
$$P_{0}\simeq 2.2 \times 10^{-9} \left( \frac{2}{5 \Omega_m}\right)^2
H_{0}^{-4} k_{\rm WMAP}^{1-n}\;,$$
where $k_{\rm WMAP}=0.02$ Mpc$^{-1}$ is a characteristic wavelength, given in 
\cite{komatsu08}.
Note that in this approach all the
mass is locked inside halos, according to the normalization
constraint: \be \int_{-\infty}^{+\infty} f_{\rm PSc}(\sigma) d{\rm
\ln}\sigma^{-1} = 1. \ee Although the above (Press-Schecther)
formulation was shown to give a reasonable approximation to the
expectations provided by numerical simulations, it was later found
to over-predict/under-predict the number of low/high mass halos at the
present epoch \cite{Jenk01,LM07}.

\begin{figure}[ht]
\mbox{\epsfxsize=8.5cm \epsffile{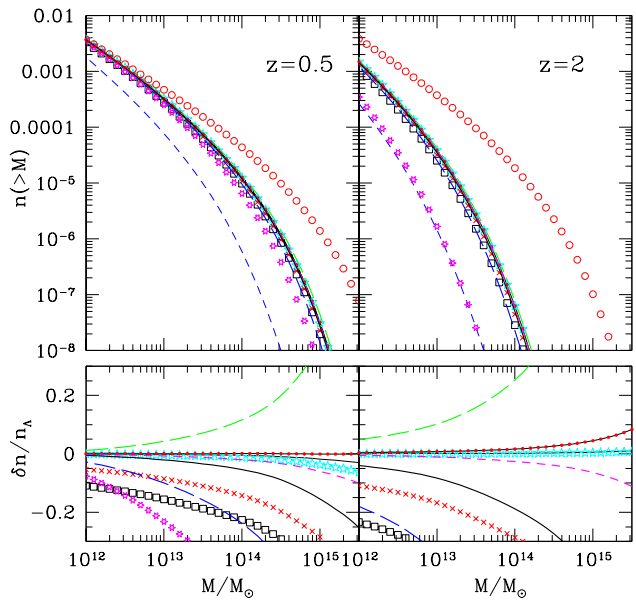}} \caption{The halo mass
function at two different redshifts. The different DE models are
represented by different symbols and/or line types (see Table 1 
for definitions).}
\end{figure}

More recently,  a large number of works have provided better fitting
functions of $f(\sigma)$, some of them  based on a phenomenological
approach. In the present
treatment, we adopt the one proposed by  Reed et al. \cite{Reed}. 

\subsection{Collapse Threshold and Mass Variance of DE models}
In order to compare the mass function predictions of the 
different DE models, it is imperative to use
for each model the corresponding values of $\delta_c$ and $\sigma^{2}(M,z)$.
In the Press-Schecther formalism, 
the rms matter fluctuations is parametrized at redshift $z=0$  so that one has for any
cosmological model:
\be
\sigma^2(M,z)=\sigma^2_8(z) \frac{\Psi(\Omega_m, R)}{\Psi(\Omega_m,
  R_8)}
\ee
with
\be \Psi(\Omega_m, R)=\int_{0}^{\infty} k^{n+2} T^{2}(\Omega_m, k) W^2(kR) dk
\ee
and
\be
\sigma_8(z)=\sigma_8(0) \frac{D(z)}{D(0)} \;,
\ee
where $\sigma_{8}(0) [\equiv \sigma_8]$ the rms mass fluctuation on $R_{8}=8 h^{-1}$ Mpc
scales at redshift $z=0$. Therefore, for a given  perturbed DE model, it 
is not enough to consider its growing mode, $D(z)$, but one has also
to obtain the corresponding $\sigma_8$ value. Below the
parameters $D_{\rm DE}(z), \sigma_{\rm 8, DE}$ denote the growth
factor and the rms mass perturbation normalization of the DE models,
while $D_{\Lambda}(z), \sigma_{8, \Lambda}$ denote the corresponding
quantities for the reference $\Lambda$CDM model.

\subsubsection{The relevant DE model $\delta_c$ values}
It is well known that in the conventional $\Lambda$ cosmology $\delta_{c} \simeq 1.675$, while Weinberg \& Kamionkowski \cite{Wein03}, provide 
an accurate fitting formula to estimate $\delta_{c}$ for any DE model 
with a constant equation of state parameter (see their eq.18). 
Let us now discuss the values of the 
linearly extrapolated density $\delta_c$ adopted here  for
the different DE models.

Firstly, since for the $\Lambda_{RG}$, $\Lambda_{PS}$ and CCDM models,
the EoS parameter is strictly equal to $-1$, 
we are completely justified to use 
$\delta_{c} \simeq 1.675$. 
Using literature data we also find: (i) for the BRG model 
$\delta_{c} \simeq 1.667$ \cite{Schmidt10}, 
(ii) for the EDE model, $\delta_{c} \simeq 1.672$
\cite{Pace10}, (iii)
for the CPL model, $\delta_{c} \simeq 1.663$ \cite{Pace10}, and (iv)
for the VCG model, $\delta_{c} \simeq 1.642$ \cite{Pace10}.
Interestingly, one can check that the above $\delta_c$ values 
can be well approximated using the previously discussed fitting formula 
\cite{Wein03}, despite the fact that 
it was derived for a constant equation of state parameter. 
As an example, in the case of the 
BRG geometrical model the fitting formula predicts
$\delta_{c} \simeq 1.667$,
which is in excellent agreement with that found by the theoretical
analysis of Schmidt et al. \cite{Schmidt10}.

Secondly, for the rest of the dark energy models (LRDE, TCM, PNGB and HRDE)
there are, to our knowledge, no available
$\delta_{c}$ values in the literature, which implies that 
one has to study in detail the spherical collapse model in order to estimate 
their exact $\delta_{c}$ values, something which is beyond the 
scope of the present paper. 
However, since in these models $w \simeq -1$ 
close to the present epoch, we have decided to adopt the 
the Weinberg \& Kamionkowski formula because it appears to 
work quite well. In this case, for the remaining 
DE models (LRDE, TCM, PNGB and HRDE)  
we have derived $\delta_{c} \simeq 1.674$.

\subsubsection{Estimation of the different DE model $\sigma_8$}
The different DE models $\sigma_8$ value can be estimated by
scaling the present time ($z=0$) $\sigma_{8, \Lambda}$ value 
to that relevant to each DE model by
using eq.(\ref{sig88}) at the present time. We again have that:
\begin{equation}\label{s88}
\sigma_{\rm 8, DE}=\sigma_{8, \Lambda} 
\frac{D_{\rm DE}(0)}{D_{\Lambda}(0)}
\sqrt{ \frac{P_{\rm DE, 0}}{P_{\rm \Lambda, 0}}
\frac{\Psi(\Omega_{m}, R_8)}{\Psi(\Omega_{m, \Lambda}, R_8)}}
\end{equation}
with $\Omega_{m,\Lambda}$ denoting the value 
for the reference $\Lambda$ model (in our case $\Omega_{m,\Lambda}=0.28$,
see section 3), while we have (for fixed $H_0$, as in our case), that:
\be
\left(\frac{P_{\rm DE, 0}}{P_{\Lambda, 0}}\right)^{1/2} = \frac{\Omega_{m, \Lambda}}{\Omega_m}\;.
\ee
It thus follows that if the $\sigma_{8,
  \Lambda}$ value is known, one  may  derive the corresponding
$\sigma_{\rm 8, DE}$ values to the other DE models.
In  this concern, the combined SNIa+BAO+WMAP5 analysis of 
Komatsu et al. \cite{komatsu08} (see also \cite{Dun09}) 
provided a value of $\sigma_{8, \Lambda}\simeq 0.812$, 
while the corresponding WMAP7 analysis yielded
[for $w(z)=-1$]: $\sigma_{8, \Lambda}\simeq 0.803$ (using the 
WMAP7 alone) and $\sigma_{8}\simeq 0.807$ for the joint
WMAP7+BAO+$H_{0}$ analysis \cite{komatsu08}. 
A recent analysis based on cluster
abundances have also furnished the following degenerate combination: 
$\sigma_{8, \Lambda}= 0.83 \pm 0.03
(\Omega_{m, \Lambda}/0.25)^{-0.41}$ \cite{Rozo09},
which for our case ($\Omega_{m,\Lambda}=0.28$) implies $\sigma_{8, \Lambda}\simeq 0.792$.
Fu et al. \cite{Fu08} 
based on a weak-lensing procedure found $\sigma_{8, \Lambda}=
0.837\pm 0.084(\Omega_{m,\Lambda}/0.25)^{-0.53}$ which implies 
$\sigma_{8, \Lambda}\simeq 0.788$ in our case.
In addition, studies  based on the peculiar velocities
statistical analysis \cite{Pike05}
obtained $\sigma_{8, \Lambda}=0.88\pm 0.05 (\Omega_{m,\Lambda}/0.25)^{-0.53}$
or $\sigma_{8, \Lambda} \simeq 0.829$ in our case. It should be stressed that  
the the average scattering of the four independent $\sigma_{8, \Lambda}$
values (of those based on WMAP, we use only the joint WMAP7+BAO+$H_{0}$ result)
is quite small ($\langle \sigma_{8, \Lambda} \rangle \simeq 0.804 \pm 0.018$),
thereby reinforcing the consistency of the different measurements. 
Finally, by inserting the latter value in Eq.(\ref{s88}) we can estimate 
the corresponding $\sigma_{\rm 8, DE}$ values, listed in Table 1, to be 
used in our mass function analysis.

For completeness, it should be remarked that 
some recent analyzes are suggesting significant higher values of
$\sigma_{8}$.  For example, 
Watkins et al. \cite{Wat09} studying the bulk flow on
scales of $\sim 100 \; h^{-1}$Mpc found a $\sigma_{8}$ normalization 
which is increased  by a factor of $\sim 2$ with  respect to the one of $\Lambda$CDM model.
If these results are correct, the
$\Lambda$CDM model would be strongly challenged. However, such a discussion is beyond the scope of
the current work.

\subsection{Halo Mass Function \& Number Counts of DE Models}
We now pass to our results. In the upper panels of Fig.\,1, for
two different redshifts ($z=0.5$ and 2), we display the integral halo mass
function, $n(>M)$, for all DE models previously discussed.  The different models
are characterized by the symbols and line-types presented in Table
1.

In the corresponding lower panels one may see the fractional
difference between each DE model with the concordance $\Lambda$CDM
model, that is, $\delta
n/n_{\Lambda}=(n_{\rm DE}-n_{\Lambda})/n_{\Lambda}$. We stress that
we have shown the case $z=2$  only for comparison of the model
expectations. In other words, it is not a statement of the viability
to actually  observe clusters at such a large redshift (not only due
to technical limitations but also because the majority of dark
matter halos are not expected to have virialized at such a
redshift, thereby producing their X-ray signatures).

It is worth noticing  the mass
function expectation of the $\Lambda_{PS}$ model (open red circles),
the BRG model (blue dashed line) and, at high redshifts, of
the CCDM model (magenta stars) 
appear to be completely different from the reference $\Lambda$CDM cosmology.

Given the halo mass function we can now derive an observable
quantity which is the redshift distribution of clusters, ${\cal
N}(z)$, within some determined mass range, say $M_1\le
M/M_{\odot}\le M_2$. This can be estimated by integrating, in mass,
the expected differential halo mass function, $n(M,z)$, according
to:

\be {\cal
N}(z)=\frac{dV}{dz}\;\int_{M_{1}}^{M_{2}} n(M,z)dM, \ee
where $dV/dz$
is the comoving volume element, which in a flat universe takes the
form:

\be \frac{dV}{dz} =4\pi r^{2}(z)\frac{dr}{dz}(z), \ee with
$r(z)$ denoting the comoving radial distance out to redshift $z$:

\be
r(z)=\frac{c}{H_{0}} \int_{0}^{z} \frac{dx}{E(x)}.
\ee

In Fig.\,2 (upper panel),  we show the
theoretically expected cluster redshift distribution, ${\cal N}(z)$,
for all the models studied (line and symbol types are listed in
Table 1), for cluster of galaxies size halos, ie., 
$M_1=10^{14} \;M_{\odot}$ and $M_2=10^{16} \;M_{\odot}$.
It is evident that many of
the different models show significant differences with respect to
the concordance $\Lambda$ model and therefore we could, in
principle, distinguish them. 
In the lower panel of Fig. 2 we show the
relative differences of the various DE models with respect to the
expectations of the concordance $\Lambda$CDM model. Note that the
three models ($\Lambda_{PS}$, CCDM and BRG models), for which we
found very large mass function differences (Fig. 1), are not shown 
because their large relative differences are beyond the limits of
the lower panel.


Let us now discuss the expected redshift distributions,
based on two future cluster surveys, and also the possibility to
observationally discriminate the different DE models.

\begin{figure}[ht]
\mbox{\epsfxsize=9cm \epsffile{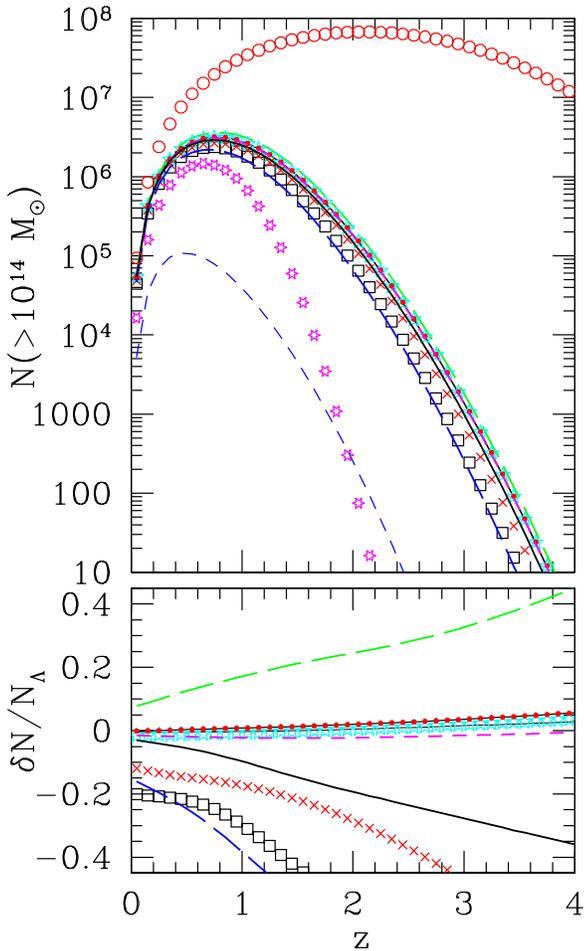}} \caption{The expected
redshift distribution of $M\magcir 10^{14} M_{\odot}$ clusters (upper
panel) of the different DE models and  the corresponding fractional
difference between the models and the reference $\Lambda$CDM
model (lower panel). The lower panel shows only those DE models that
have fractional relative differences, with respect to the $\Lambda$CDM 
model, of $\mincir 45\%$. 
Symbols correspond to the different DE models as indicated in Table 1.}
\end{figure}
These two realistic future surveys are:

\noindent (a) the {\tt eROSITA} satellite X-ray survey, with a flux
limit of: $f_{\rm lim}=3.3\times 10^{-14}$ ergs s$^{-1}$ cm$^{-2}$,
at the energy band 0.5-5 keV and covering $\sim 20000$ deg$^{2}$ of
the sky,

\noindent (b) the South Pole Telescope SZ survey, with a limiting
flux density at $\nu_0=150$ GHz of $f_{\nu_0, {\rm lim}}=5$ mJy and
a sky coverage of $\sim 4000$ deg$^{2}$.

\begin{figure}[ht]
\mbox{\epsfxsize=8.8cm \epsffile{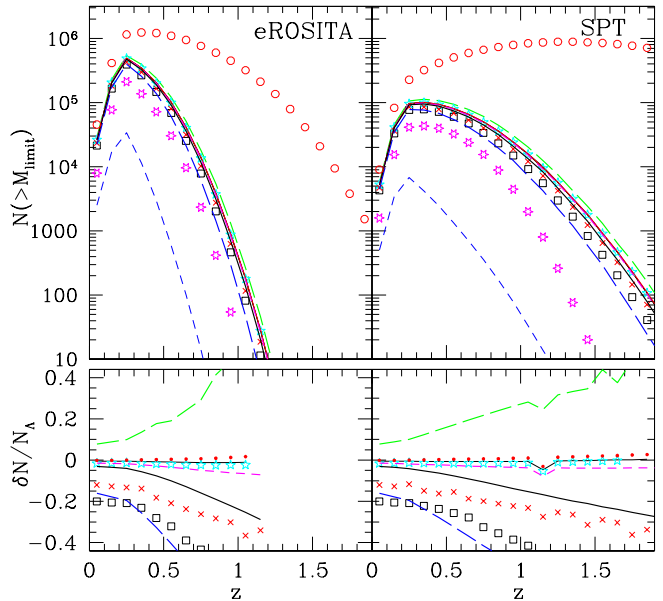}} \caption{The expected
cluster redshift distribution of the different DE models and for the 
two different future cluster
surveys (upper panels), and the corresponding fractional difference
with respect to the reference $\Lambda$CDM model (lower
panels). Symbols correspond to the different DE models as indicated in
Table 1.}
\end{figure}

To realize the predictions of the first survey we use the relation
between halo mass and bolometric X-ray luminosity, as a function of
redshift, provided in \cite{Fedeli}, ie: \be\label{bolom}
L(M,z)=3.087 \times 10^{44} \left[\frac{M E(z)}{10^{15} h^{-1}
    M_{\odot}} \right]^{1.554} h^{-2} \; {\rm erg s^{-1}} \;.
\ee The limiting halo mass that can be observed at redshift $z$ is
then found by inserting in the above equation the limiting
luminosity, given by: $L=4 \pi d_L^2 f_{\rm lim}${\em c}$_b$, with
$d_L$ the luminosity distance corresponding to the redshift $z$ and
{\em c}$_b$ the band correction, necessary to convert the bolometric
luminosity of eq.(\ref{bolom}) to the 0.5-5 keV band of {\tt
eROSITA}. We estimate this correction by assuming a Raymond-Smith
(1977) plasma model with a metallicity of 0.4$Z_{\odot}$, a typical
cluster temperature of $\sim 4$ keV and a Galactic absorption column
density of $n_{H}=10^{21}$ cm$^{-2}$.

The predictions of the second survey can be realized using again the
relation between limiting flux and halo mass from \cite{Fedeli}:
\be\label{sz} f_{\nu_0, {\rm lim}}= \frac{2.592 \times 10^{8} {\rm
mJy}}{d_{A}^{2}(z)} \left(\frac{M}{10^{15} M_{\odot}}\right)^{1.876}
E^{2/3}(z) \; \ee where $d_A(z) \equiv d_L/(1+z)^2$ is the angular
diameter distance out to redshift $z$.

In Fig.\,3 (upper panel) we present the expected redshift
distributions above a limiting halo mass, which is $M_1 \equiv M_{\rm
limit}=\max[10^{14} M_{\odot}, M_f]$, with $M_f$ corresponding to
the mass related to the flux-limit at the different redshifts,
estimated by solving eq.(\ref{bolom}) and eq.(\ref{sz}) for $M$. In
the lower panels we present the fractional difference between the
different DE models and $\Lambda$CDM, similarly to Fig.\,2,
but now for the realistic case of the previously mentioned future cluster
surveys. It is evident that the imposed flux-limits together with
the scarcity of high-mass halos at large redshifts, induces an
abrupt decline of ${\cal N}(z)$ with $z$, especially in the case
of the {\tt eROSITA} X-ray survey (note the shallower redshifts depicted in
Fig.3 with respect to Fig.2).

In  the lower panels of Fig. 3 we display the relative differences of
 the DE models but only up to a redshift at which they are
significant, that is, such that:
\be\label{stat} \frac{{\cal
N}_{\rm DE}-{\cal N}_{\Lambda}}{ ({\cal N}_{\rm DE}-{\cal
N}_{\Lambda})^{1/2} }> 3.5,
\ee
with ${\cal N}_{\rm DE}$ the redshift
distribution predicted by some DE model and ${\cal
N}_{\Lambda}$ the corresponding redshift distribution of
$\Lambda$CDM model. However, such a criterion does not take into
account cosmic variance and possible observational systematic
uncertainties which can hamper detecting small (but according to
Eq. \ref{stat} significant) relative differences. We believe that
relative differences of $\mincir 5\%$ will be difficult to detect
especially at relative high redshifts.

In Table 1, one may see a more compact presentation
of our results including the
relative fractional difference between all DE models and the 
$\Lambda$CDM model, in 3 distinct  redshift bins and for both
future surveys.

Based on our ${\cal N}(z)$ analysis and the results presented in
Figure 3 and Table 1, we can now divide the studied
DE models into those that can be distinguished observationally
and those that are practically indistinguishable from the current
paradigm ($\Lambda$CDM). The latter DE models 
are the following four: XCDM, HRDE, TCM and PNGB. One has to remember
however that these results are based on using DE model parameters that
have been fitted by the present day cosmological data (see section
4). As an illustrative
example, we remind the reader that the XCDM model, compared here,
is one with $w=-0.99$; if future
cosmological data would provide a different value for the equation of
state parameter then the $N(z)$ predictions of such an XCDM model could be quite
different than those derived here.

Regarding the models that are distinguishable with respect to the
concordance model, three of them (BRG, $\Lambda_{PS}$ and CCDM) show
extremely large variations making it trivial to distinguish them.
From the rest of the distinguishable DE models all of them show clear
signs of difference, at all redshifts and in both future cluster surveys, 
with respect to the $\Lambda$CDM expectations;
only the $\Lambda_{RG}$ model needs to be
distinguished at higher redshifts ($z\magcir 0.5$).

As an additional test and in order to check the sensitivity of our results only on the functional
form of the DE equation of state parameter, we have imposed a unique value
of the rms mass fluctuation normalization to all DE models
according to: $\sigma_{8, DE}=\sigma_{8, \Lambda}$, and repeated the cluster-size halo $N(z)$
analysis. We now find slightly different results although
in the same overall direction with our main analysis. For example, 
the DE models that cannot be distinguished 
from the reference $\Lambda$CDM model are now the XCDM, TCM, PNGB, EDE and
$\Lambda_{RG}$, with the first three being common in both
analyzes, the $\Lambda_{RG}$ model showing a small difference and 
the EDE being the only model that shows a significantly different
behavior between the two analyzes.


Finally we would like to mention that an interesting paper 
appeared recently \cite{suman10} and  
among other issues, it compares different forms
of the halo mass function and its redshift evolution using N-body
simulations of the $\Lambda$CDM and $w$CDM
($w=$const) models. They do find some differences at the few percent level.
Although our analysis is
self-consistent, in the sense that we compare the expectations of DE
models with respect to those of the concordance cosmology using the
same mass function model, we plan
to investigate in a forthcoming paper how sensitive are our
observational predictions 
to the different mass functions fitting formulas.

\section{Concluding Remarks}

In this paper,  we have investigated the cluster abundances beyond
the conventional $\Lambda$CDM cosmology by using several
parameterizations for the dark energy.  In order to do that,  we
first performed a joint likelihood analysis using the most
recent high quality cosmological data (SNIa, CMB shift
parameter and BAOs), thereby obtaining tight
constraints on the main cosmological and dark energy free parameters.
At the level of the resulting Hubble function,
we have found that of all dark energy models  (apart a Brane world
cosmology), are statistically indistinguishable (within 1$\sigma$) from a
flat $\Lambda$CDM model, as long as they are confronted with the
quoted set of observations.

On the other hand, despite the fact that these models
closely reproduce the $\Lambda$CDM Hubble expansion, 
we show that eight out of the twelve studied DE models, using
the observationally fitted model parameters,
can be differentiated from the reference $\Lambda$CDM model on the
basis of their redshift distribution of cluster-size halos.
Such a comparison was made possible by using the expectations of a
future X-ray (based on the {\tt eROSITA}  Satellite) and SZ
cluster surveys (based on the South Pole Telescope).

The main comparison results 
can be summarized in the following statements (for nomenclature see section 4):
\begin{itemize}
\item Four DE models, namely, XCDM, HRDE, TCM and PNGB
cannot be distinguished from the $\Lambda$CDM model at any significant level.
\item Seven models, ie., the BRG, CPL, LRDE, EDE, VCG, $\Lambda_{PS}$, CCDM, can be easily 
distinguished due to the fact that they show strong and significant variations with
respect to the concordance $\Lambda$ model even at $z=0$, implying
that even with the present day surveys one could effectively
distinguish them.
\item The $\Lambda_{RG}$ model, although presents 
  relatively small variations with respect to the concordance $\Lambda$ model, 
 it can be clearly distinguished at relatively high redshifts ($z\magcir 0.5$).
\end{itemize}

In a future work we will present a comparison between our model
predictions and the observationally determined cluster mass function
at different redshifts as well as the available (X-ray or optical)
cluster redshift distribution.

\vspace {0.4cm}

\acknowledgments The authors thank the anonymous referee for useful 
comments and suggestions. MP acknowledges funding by
Mexican CONACyT grant 2005-49878, and JASL is partially supported by
CNPq and FAPESP under grants 304792/2003-9 and 04/13668-0,
respectively.

\end{document}